\documentclass[reprint,aps,nofootinbib,superscriptaddress,preprintnumbers]{revtex4-2}
\usepackage{comment}
\usepackage{tikz}
\usetikzlibrary{arrows.meta,calc}
\usetikzlibrary{decorations.pathmorphing,shapes,positioning}
\usepackage[bottom]{footmisc}
\usepackage{graphicx} 

\usepackage{braket}
\usepackage{amsmath,amssymb,amsfonts}

\usepackage{dcolumn}                
\usepackage{bm}                     
\usepackage{ulem}

\usepackage{mathtools}
\usepackage[dvipsnames]{xcolor}

\newcommand\blfootnote[1]{%
  \begingroup
  \renewcommand\thefootnote{}\footnote{#1}%
  \addtocounter{footnote}{-1}%
  \endgroup
}

\begin{document}

\preprint{CALT-TH 2025-037}

\title{Composite AdS geodesics for CFT correlators and timelike entanglement entropy}

\author{Hardik Bohra}
\affiliation{Department of Physics and Astronomy, University of Kentucky, Lexington, KY 40506}

\author{Allic Sivaramakrishnan}
\affiliation{Walter Burke Institute for Theoretical Physics, California Institute of Technology, Pasadena, CA 91125}

\blfootnote{bohra.hardik@uky.edu, allic@caltech.edu}

\begin{abstract}

We study how to recover timelike worldlines in AdS from CFT data as a toy model for holographically reconstructing realistic observers. We give a bulk extremization procedure that determines composite timelike-spacelike geodesics that connect timelike-separated boundary points. The total geodesic length matches the length extracted from CFT correlators at the timelike-separated points. We show agreement in Poincaré AdS, for generic boundary points in global AdS, and also for the BTZ solution, in which the timelike segment probes behind the horizon. We refine related methods to compute timelike entanglement entropy in AdS$_3$/CFT$_2$ and recover known results.

\end{abstract}

\maketitle

\section{Introduction}

Holography offers a well-controlled way to study observers and observables in quantum gravity. Approaches to bulk reconstruction focus on recovering field operators in anti-de Sitter space (AdS) from the dual conformal field theory (CFT). Methods include HKLL reconstruction \cite{Hamilton:2006az, Kabat:2011rz}, ideas arising from the Ryu-Takayanagi formula \cite{Ryu:2006bv, Harlow:2018fse}, and work on recovering operators within the black hole interior \cite{Papadodimas:2012aq, Papadodimas:2013jku, Akers:2022qdl}. Graviton contributions are challenging to include in these approaches, but are more easily incorporated if we model a realistic observer by its worldline, and measurements by operator insertions along the worldline (see \cite{Giddings:2025xym} for a review).

CFT correlation functions capture bulk effective field theory \cite{Gary:2009ae, Polchinski:1999yd, Heemskerk:2009pn, Penedones:2010ue, Okuda:2010ym, Maldacena:2015iua, Aharony:2016dwx, Meltzer:2019nbs, Meltzer:2020qbr, Caron-Huot:2025she, Caron-Huot:2025hmk} and can probe the black hole interior \cite{Festuccia:2005pi, Hubeny:2006yu, Festuccia:2008zx, Kraus:2002iv, Papadodimas:2012aq, Ceplak:2024bja, Grinberg:2020fdj, Afkhami-Jeddi:2025wra, Ceplak:2025dds, Dodelson:2025jff} via, in the simplest settings, the geodesic or WKB approximation: the Euclidean vacuum two-point function of a scalar $\phi$ with mass $m$ behaves as $\braket{\phi(x_1)\phi(x_2)} \sim e^{-m \ell(x_1,x_2)}$ for $m \ell(x_1,x_2) \gg 1$, with geodesic distance $\ell(x_1,x_2)$. In this regime, CFT correlators at spacelike separations are computed by boundary-anchored spacelike geodesics. Whether something similar happens for timelike separations is less clear, as AdS does not have boundary-anchored timelike geodesics (but see \cite{Balasubramanian:2012tu, Arefeva:2016nic, He:2024emd, Czech:2016tqr, Czech:2016xec, Sarosi:2017rsq, Haehl:2025ehf}).

However, timelike entanglement entropy \cite{Doi:2022iyj, Doi:2023zaf, Milekhin:2025ycm} in the CFT is computed in AdS by composite codimension-two extremal surfaces, and in AdS$_3$/CFT$_2$ these are chains of timelike and spacelike geodesics that connect two timelike-separated boundary points. The geometry and surfaces are sometimes complexified \cite{Doi:2022iyj, Doi:2023zaf, Heller:2024whi, Heller:2025kvp}, raising questions about extensions to non-analytic setups.

Here, we give a prescription for determining composite, real geodesics that connect timelike-separated boundary points $x_1,x_2$ in AdS$_{d+1}$/CFT$_d$ in the original geometry:
\begin{enumerate}
    \item Compute the total proper length $\ell_s$ of geodesics that connect boundary points $x_1,x_2$ to bulk points $y_1,y_2$ respectively, and the proper time $\ell_t$ of a geodesic between $y_1, y_2$.
    \item Extremize $\ell_s, \ell_t$ over $y_1,y_2$.
    \item To fix any remaining degeneracy, compare $\ell_s+i\ell_t$ to $\ell(x_1,x_2)$ obtained by analytically continuing $x_1, x_2$ to timelike separations with a specific $i\epsilon$ prescription that we give.
\end{enumerate}

\begin{figure}[ht]
  \centering
\includegraphics[width=0.40\textwidth]{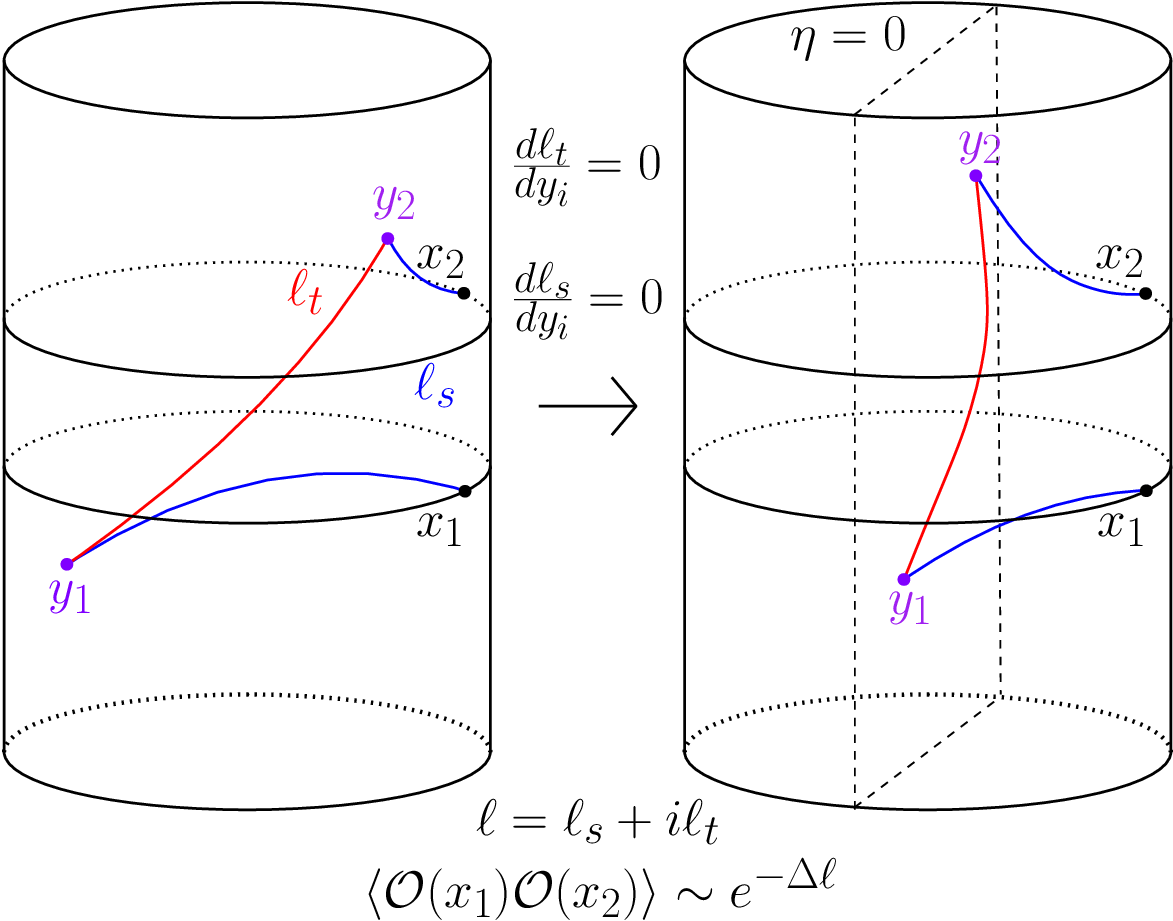}
  \caption{For timelike-separated boundary points $x_1, x_2$, we extremize the lengths $\ell_s, \ell_t$ over bulk points $y_1, y_2$. We show this uniquely specifies a composite geodesic whose complex length $\ell$ is equal to the length extracted from $\braket{\mathcal{O}(x_1) \mathcal{O}(x_2)}$.}
  \label{fig:stacked}
\end{figure}

We find that this prescription, depicted in Figure \eqref{fig:stacked}, determines composite geodesics with a unique complex length $\ell(x_1,x_2)$. We further show that $\ell(x_1,x_2)$ matches the length extracted from time-ordered boundary two-point functions $\braket{\mathcal{O}(x_1) \mathcal{O}(x_2)}$. 

Our results reproduce timelike entanglement entropy in AdS$_3$/CFT$_2$. Our prescription refines the approaches in \cite{Doi:2022iyj, Doi:2023zaf,Heller:2024whi, Heller:2025kvp} by using only the original, uncomplexified geometry and its geodesics, and by making no assumptions about the points that join the segments.

We compute $\ell(x_1,x_2)$ in Poincaré AdS in Section \ref{poincare}, in global AdS for generic $x_1, x_2$ in Section \ref{global}, and in the BTZ solution in Section \ref{BTZ}. We confirm our results by studying the geodesics themselves in Appendices \ref{sec:GlobalGeodesicMethod} and \ref{sec:BTZB}. In Appendix \ref{modularobservers}, we review how proper time differs from modular time.

\section{Poincare AdS}
\label{poincare}
AdS admits geodesics that connect boundary points $x_1, x_2$ when they are spacelike separated, but not when timelike separated. However, working in Poincar\'e AdS, we will analyze solutions to the geodesic equation and find that timelike-separated boundary points $x_1, x_2$ can indeed be connected by a composite geodesic---a curve composed of concatenated timelike and spacelike geodesic segments. For definiteness, we consider AdS$_3$/CFT$_2$, although our results apply trivially to AdS$_{d+1}$/CFT$_d$. The metric is 
\begin{equation}
    ds^2 = \frac{1}{z^2}(-dt^2 + dw^2 + dz^2),
\end{equation}
where $z=0$ is the AdS boundary, and we set $R_{AdS} = 1$. The boundary points are $x_i = (t_i,w_i)$ and the bulk points $y_i = (t_i,w_i,z_i)$. 

We first consider the spacelike configuration $(x_1-x_2)^2 > 0 $. We parameterize bulk curves between $x_1$ and $x_2$ by an affine parameter $\lambda$ normalized such that the velocity $\dot{y}^\mu \equiv (\dot{t},\dot{w},\dot{z})$ obeys $\dot{y}^\mu \dot{y}_\mu= 1$. The geodesic equations are
\begin{equation}\label{eqn:PoincareGeodesicEq}
    \dot{t} = Ez^2, \qquad \dot{w} = Pz^2, \quad \dot{z}^2 = z^2 \big(1 - Q^2z^2 \big),
\end{equation}
where $E, P$ and $Q^2 = -E^2 +P^2$ are constants of motion determined by boundary data, $ Q^{-2} = \frac{1}{4}(x_1 - x_2)^2$. Equation \eqref{eqn:PoincareGeodesicEq} shows that the geodesic, which is a semicircle, has a turning point in $z$. We can solve \eqref{eqn:PoincareGeodesicEq} with the AdS boundary located at $\lambda = \pm \infty$.

To compute the geodesic length $\ell(x_1,x_2)$, we introduce a cutoff surface at $z = \delta \ll 1$. The boundary is then located at $\pm \lambda_{\max}$ and we recover the familiar length formula,
$
    \ell = 2\lambda_{\max} = \log\left( (x_1 - x_2)^2/\delta^2\right).
$

Now, we analytically continue $\ell$ in $x_1, x_2$ to reach the timelike configuration $(x_1-x_2)^2 < 0 $. For reasons that will become clear, of the two possible continuations, we choose the one in which $\ell$ acquires a positive imaginary contribution,
\begin{equation}
    \ell = i\pi + 2\log\left( \sqrt{(t_1-t_2)^2 - (w_1-w_2)^2}/\delta\right).
\end{equation}
Whether this $\ell$ is the length of some curve is now unclear.

We can in fact show that $\ell = i \ell_t + \ell_s$ is the length of a bulk curve connecting $x_1$ and $x_2$ that is a union of a timelike geodesic with length $\ell_t$ and two spacelike geodesics with total length $\ell_s$. For timelike-separated boundary points, the boundary-anchored spacelike segments described by equation \eqref{eqn:PoincareGeodesicEq} have $Q^2<0$, which implies that $z(\lambda)$ has no turning point. The spacelike geodesics anchored at $x_1$ and $x_2$ therefore originate at $z = \delta$ and reach the Poincar\'e horizons, $z = \infty$ with $t = \pm \infty$. This was found in \cite{Doi:2022iyj, Doi:2023zaf}.

We can further identify a timelike geodesic between the horizons. Setting $w_1 = w_2$ for simplicity, a timelike geodesic in the Poincaré patch parameterized by $\tau \in [-\pi/2,\pi/2]$ with $\dot y^\mu \dot y_\mu=-1$ satisfies 
$-\dot{t}^2+\dot{z}^2 = -z^2$ and $\dot{t} = p_t z^2$. 
Its solution $\big(t(\tau), w_1, z(\tau)\big)$ satisfies $z^2 - (t-t_0)^2 = p_t^{-2}$ for constants $t_0$ and $p_t$. From this, we can see that timelike geodesics cannot reach the boundary for finite $p_t$: as $\dot{z}^2 = z^2(p_t^2 z^2-1)$, the region $z < 1/p_t$ is forbidden. 

The length $\ell_t$ of this geodesic is 
\begin{align}
        \ell_t &= 2 \int_{1/p_t}^\infty \frac{dz}{z\sqrt{p_t^2z^2-1}} = \pi.
\end{align}
The same result holds for $w_2 - w_1 \neq 0$. The length is independent of $p_t$, and so there is a family of distinct geodesics parameterized by $p_t$, whose value does not appear to be fixed by boundary conditions. 

For comparison, the time-ordered CFT two-point function of two primaries is $\braket{0|\mathcal{O}(x_1) \mathcal{O}(x_2)|0} = 1/((x_1-x_2)^2+i\epsilon)^\Delta$, and in the spacelike configuration, 
\begin{equation}
\ell = -\frac{1}{\Delta}\log \braket{0|\mathcal{O}(x_1) \mathcal{O}(x_2)|0} = \log((x_1-x_2)^2+i\epsilon).
\end{equation}
Analytically continuing to timelike separations gives
\begin{equation}
\ell = \log(|x_1-x_2|^2)+i\pi = \ell_s + i \ell_t,
\end{equation}
in agreement with the geodesic calculation we provided. Note that the time-ordered two-point function corresponds to a specific $i\epsilon$ prescription that gives $i\pi$ regardless of the sign of $t_1-t_2$. This choice of $i\epsilon$ specifies the analytic continuation of the length, which we will take as the definition of this continuation from now on. 

\section{Global AdS}
\label{global}
In global AdS coordinates, the extremization prescription summarized in the introduction produces a unique piecewise-geodesic curve between $x_1, x_2$. We work in global AdS$_3$, but our results apply without loss of generality to AdS$_{d+1}$. The metric is
\begin{equation}
ds^2 = - \cosh^2\!\rho\ dt^2 + d\rho^2 + \sinh^2\!\rho \ d\phi^2,
\end{equation}
where $\rho \geq 0$ and the AdS boundary is located at $\rho=\infty$. The geodesic distance between bulk points in AdS can be expressed in embedding-space coordinates. For the embedding space $\mathbb{R}^{2,2}$, with metric $\text{diag}(-1,-1,1,1)$, the AdS geodesic distance between points $X_I$ and $X_J$ is 
\begin{equation}\label{eqn:embeddistance}
    \cosh{L} = X^0_I X^0_J + X^1_I X^1_J - X^2_I X^2_J - X^3_I X^3_J.
\end{equation}
To exploit the symmetries of the setup, we change coordinates to an AdS$_2$ slicing of global AdS$_3$ (see, for example \cite{Doi:2023zaf}),
\begin{equation}
     ds^2 = d\eta^2 + \cosh^2(\eta) \big(-\cosh^2(r) dt^2 + dr^2 \big),
\end{equation}
with $-\infty < \eta, r < \infty$, and where $\eta = \pm \infty$ is the boundary. The relation to embedding space is 
\begin{equation}\label{eqn:AdS2slice}
    \begin{split}
        X^0 &= \cosh(\eta) \cosh(r) \sin(t), \ 
        X^2 = \sinh(\eta), \\
        X^1 &= \cosh(\eta) \cosh(r) \cos(t), \
        X^3 =\cosh(\eta) \sinh(r).
    \end{split}
\end{equation}
Following \eqref{eqn:embeddistance} and \eqref{eqn:AdS2slice}, the geodesic distance is 
\begin{align}
      \cosh{L} = &\cosh{\eta_1} \cosh{\eta_2}  \big(\cosh{r_1}\cosh{r_2} \cos(t_1 - t_2) 
      \nonumber
      \\
      &~~~~~- \sinh{r_1} \sinh{r_2}\big)- \sinh{\eta_1} \sinh{\eta_2}. \label{eqn:Geodesicdistance}
\end{align}
The geodesic distance between boundary points $x_1=(T_1, R_1)$ and $x_2= (T_2, R_2)$ is
\begin{align}\label{eqn:BdyGeodesic_length}
     \ell = \log\!\left[ \frac{1}{2\delta^2} \Big( - 1 + \cosh\!{R_1}\cosh\!{R_2} \cos(T_1 - T_2) \right.\nonumber\\ \left. - \sinh\!{R_1} \sinh\!{R_2}\Big) \right],
\end{align}
with radial cutoff $\delta = \exp(-\eta_\infty)$. The boundary is AdS$_2$, which implies that the boundary light cone is defined by where the logarithm's argument in \eqref{eqn:BdyGeodesic_length} vanishes. Analytically continuing $\cosh L$ from the spacelike to timelike configuration therefore gives
\begin{align}\label{eqn:t-like_length}
    \ell = i\pi +\log\!\Big[ \Big( &1 - \cosh{R_1}\cosh{R_2} \cos(T_1 - T_2) 
    \nonumber
    \\
    &
    + \sinh{R_1} \sinh{R_2}\Big)/(2\delta^2)  \bigg].
\end{align}
This agrees with the length extracted in the usual way from the vacuum two-point function, $\ell = -(1/\Delta)\log \braket{\mathcal{O}(R_1,T_1) \mathcal{O}(R_2,T_2)}$ in global coordinates at timelike separations.\footnote{To see this, note that in embedding space, $\braket{\mathcal{O}(P_1) \mathcal{O}(P_2)} = 1/(-P_1 \cdot P_2)^{\Delta}$ with projective boundary points $P_i=\big(\cosh\! R_i\,\sin T_i,\ \cosh\! R_i\,\cos T_i,\ 1,\ \sinh\!R_i\big)$ reproduces \eqref{eqn:BdyGeodesic_length}. To obtain the two-point function in the standard cylindrical coordinates, one instead uses the parametrization $P(t,\phi)=(\sin t,\ \cos t,\ \cos\phi,\ \sin\phi)$.}

We now apply our prescription to identify a boundary-anchored bulk curve whose length is \eqref{eqn:t-like_length}.
We first focus on a bulk-to-bulk timelike geodesic. Denoting the endpoints that solve the extremization conditions as 
$(\bm{\eta}_1, \bm{t}_1, \bm{r}_1)$ and $(\bm{\eta}_2, \bm{t}_2, \bm{r}_2)$, we extremize \eqref{eqn:Geodesicdistance},
\begin{equation}\label{eqn:TimelikeCondition}
    \frac{dL}{d\eta_1} = 
    \frac{dL}{d\eta_2} = 
    \frac{dL}{dt_1} = 
    \frac{dL}{dr_1} =
    \frac{dL}{dt_2} =
    \frac{dL}{dr_2} =0.
\end{equation}
Solving $\frac{dL}{d\eta_i}=0$ gives $\bm{\eta}_1 = \bm{\eta}_2 = 0$, for which

\begin{align}
&\frac{dL}{dr_1} = 
\sinh{r_1}\, \cosh{r_2}\, \cos(t_1 - t_2) - \cosh{r_1}\, \sinh{r_2}, 
\nonumber
\\
&\frac{dL}{dt_1} =
-\, \cosh{r_1}\, \cosh{r_2}\, \sin(t_1 - t_2) ,
\nonumber
\\ 
&\frac{dL}{dr_2} =
\cosh{r_1}\, \sinh{r_2}\, \cos(t_1 - t_2) - \sinh{r_1}\, \cosh{r_2}, 
\nonumber
\\
&\frac{dL}{dt_2} =
 \cosh{r_1}\, \cosh{r_2}\, \sin(t_1 - t_2).
\end{align}
Solving $dL/dt_i = 0$ gives $t_1 - t_2 = n\pi$. In this case, $dL/dr_i = 0$ yields $\tanh{r_1} = (-1)^n \tanh{r_2}$. Altogether, we have  
\begin{equation}
    \bm{t}_1 - \bm{t}_2 = n \pi,\,\,\, \bm{r}_1 = (-1)^n \bm{r}_2\ , \quad n \in \mathbb{Z}.
    \label{choice_of_n}
\end{equation}
Comparison to \eqref{eqn:t-like_length} indicates we should choose the $n=\pm 1$ solution, as explained in Appendix \ref{sec:GlobalGeodesicMethod}. Without loss of generality, we choose the time-ordered configuration $n=1$, as choosing $n=-1$ is equivalent to swapping $y_1, y_2$, which gives the same geodesic upon extremization.
\begin{equation}\label{eqn:t-like_condition}
    \bm{t}_1 - \bm{t}_2 = \pi,\ ~~~~~\bm{r}_1 = -\bm{r}_2
\end{equation}
gives $L =i \pi$. The two bulk endpoints are timelike separated with proper time $\ell_t = \pi$ between them. 

Next, we extremize the total length $\ell_s$ of two spacelike segments that connect $(T_i, R_i)$ with each endpoint of the timelike segment, $(\bm{t}_i, \bm{r}_i, \bm{\eta}_i = 0)$,
\begin{multline}
    \ell_s = \log\! \Big[ \big(\cos(\bm{t}_1-T_1) \cosh\bm{r}_1\cosh\!{R_1} - \sinh{\bm{r}_1}\sinh\!{R_1} \big) \\ ~~~~\times \big(-\cos(\bm{t}_1-T_2) \cosh\bm{r}_1\cosh\!{R_2} + \sinh{\bm{r}_1}\sinh\!{R_2}\big)/\delta^2 \Big].
\end{multline}
Simplifying the expressions obtained by extremizing via $d\ell_s/d\bm{t}_1 = d\ell_s/d\bm{r}_1 = 0$ gives
\begin{gather}
    0 = \cosh^2\!{R_1}\, \sin^2(\bm{t}_1-T_1) - \cosh^2\!{R_2}\, \sin^2(\bm{t}_1-T_2),  \nonumber \\
    \tanh \bm{r}_1 = 
    \frac{\sin\!\big(2\bm{t}_1 - T_1 - T_2\big)}
       {\tanh\! R_1\,\sin(\bm{t}_1 - T_2) + \tanh\! R_2\,\sin(\bm{t}_1 - T_1)} .
\end{gather}
We have checked numerically for a variety of $(T_i, R_i)$ that there is a unique choice of $(\bm{t}_1, \bm{r}_1)$ with $\mathrm{Im} (\ell_s)=0$. In all cases, we find that $\ell_s+i \ell_t = \ell$, where $\ell$ is given in \eqref{eqn:t-like_length}. This agrees with the length extracted from the two-point function, as advertised.

A simple example is the $R_1 = R_2 = 0$ case. The extremization conditions give
    \begin{equation}
         \sin^2(\bm{t}_1 -T_1) - \sin^2(\bm{t}_1 -T_2)=0, \quad \tanh{\bm{r}_1} = 0.
    \end{equation}
The appropriate solution is
    \begin{equation}
        \bm{t}_1 = \frac{\pi}{2} + \frac{T_1+T_2}{2} ,  ~~ \bm{t}_2 = -\frac{\pi}{2} + \frac{T_1+T_2}{2} , ~~\bm{r}_1 = \bm{r}_2 = 0 .  
    \end{equation}
The special case $T_2 = -T_1$ was studied in \cite{Doi:2023zaf, Doi:2022iyj}.

In \cite{Doi:2022iyj, Doi:2023zaf}, $n=1$ was chosen and $\eta = 0$ imposed, although justification for these choices was not provided. The prescription we used derives these conditions. Our prescription is a partial refinement of \cite{Doi:2022iyj, Doi:2023zaf}, as we work in the original geometry rather than the Wick-rotated setup, we include an $i\epsilon$ prescription to specify an analytic continuation, and we compare to the analytically continued spacelike length to fix degeneracies.

The prescription we used could have failed upon extremization, as there could have been no segments with real length that solved the constraints. Its success therefore appears to be a non-trivial check. We also applied our approach for arbitrary boundary points, going beyond the specific, symmetric boundary configuration $T_2=-T_1, R_1=R_2=0$ studied in \cite{Doi:2022iyj, Doi:2023zaf}. See Appendix \ref{sec:GlobalGeodesicMethod} for an analysis of the geodesics themselves.

\section{BTZ}\label{BTZ}

We next apply our extremization procedure to the non-rotating BTZ black hole in Kruskal coordinates \((u,v,\phi)\). The relation to \(\mathrm{AdS}_3\) embedding coordinates is
\begin{equation}\label{eqn:kruskal}
\begin{split}
X^{0} &= \frac{u+v}{1+uv},\quad
X^{1} = \frac{1-uv}{1+uv}\,\cosh\!\left(r_0\,\phi\right),\\
X^{3} &= \frac{v-u}{1+uv}, \quad
X^{2} = \frac{1-uv}{1+uv}\,\sinh\!\left(r_0\,\phi\right).
\end{split}
\end{equation}
The metric is
\begin{equation}\label{eqn:BTZmetric}
    ds^2 = - \frac{4}{(1+uv)^2} du dv + \frac{(1-uv)^2}{(1+uv)^2} r_0^2 d\phi^2,
\end{equation}
where $-\infty<u,v<\infty$, and the horizon radius is $r_0$. The surfaces $uv=0$ are the future and past horizons. The conformal boundary is $uv=-1$ and the singularity is $uv=1$. Regions with $-1 \le uv <0$ lie outside the horizon, while regions with $0 < uv \le 1$ are inside.

The embedding–space distance formula \eqref{eqn:embeddistance} together with \eqref{eqn:kruskal} gives the geodesic distance between arbitrary points $(u_1,v_1,\phi_1)$ and $(u_2,v_2,\phi_2)$,
\begin{align}  
\label{eqn:coshLBTZ}
&\cosh{\!L_{\text{BTZ}}} = 
\nonumber
\\ 
&\frac{2(u_1 v_2 + u_2 v_1) + (1- u_1 v_1)(1-u_2v_2)\ \cosh\!{(r_0(\phi_2-\phi_1))}}{(1+u_1 v_1)(1 + u_2 v_2) }.
\end{align}
To obtain the spacelike geodesic between boundary points $(U_1,V_1,\Phi_1)$ and $(U_2,V_2,\Phi_2)$, we regulate boundary-anchored geodesics by $U_i V_i=-1+2\delta$ with $0<\delta \ll 1$,
\begin{equation}
\label{eqn:Lbdy}
    \ell_{\text{BTZ}} =
    \log\!\left[
    -\,\frac{V_1^{2}+V_2^{2}-2V_1V_2\cosh\!\big(r_{0}(\Phi_1-\Phi_2)\big)}
            {V_1 V_2\,\delta^2}
    \right].
\end{equation}
For purely timelike separation on the boundary ($\Phi_1=\Phi_2$),
\begin{equation}
\label{eqn:Lbdy_timelike}
    \ell_{\text{BTZ}} =\;
    i\pi\; + \;\log\!\left(\frac{(V_1-V_2)^2}{V_1 V_2\,\delta^2}\right).
\end{equation}

We now apply our extremization prescription. With bulk endpoints $(\bm u_1,\bm v_1)$ and $(\bm u_2,\bm v_2)$, extremizing the length \eqref{eqn:coshLBTZ} between timelike-separated points gives 
\begin{align}
\label{TimelikeExtremizationConditions}
    \frac{dL}{du_1} =&-\frac{2\,(1 + u_{2} v_{1})\,(v_{1} - v_{2})}{(1 + u_{1} v_{1})^{2} (1 + u_{2} v_{2})} = 0,\nonumber \\
    \frac{dL}{du_2} =& \frac{2\,(v_{1} - v_{2})\,(1 + u_{1} v_{2})}{(1 + u_{1} v_{1}) (1 + u_{2} v_{2})^{2}} = 0, \nonumber \\
    \frac{dL}{dv_1} =& -\frac{2\,(u_{1} - u_{2})\,(1 + u_{1} v_{2})}{(1 + u_{1} v_{1})^{2} (1 + u_{2} v_{2})} = 0, \nonumber \\
    \frac{dL}{dv_2} =& \frac{2\,(u_{1} - u_{2})\,(1 + u_{2} v_{1})}{(1 + u_{1} v_{1}) (1 + u_{2} v_{2})^{2}} = 0 .
\end{align}
Comparing to \eqref{eqn:Lbdy_timelike} instructs us to exclude the coincident-point solution. The conditions \eqref{TimelikeExtremizationConditions} then imply
\begin{equation}\label{eqn:t_likecondBTZ}
    1+ \bm{u}_1 \bm{v}_2 = 0, \quad 1+ \bm{u}_2 \bm{v}_1 = 0,
\end{equation}
for which \eqref{eqn:coshLBTZ} gives
$\cosh L_{\text{BTZ}}=-1$. This confirms these bulk points are timelike separated and that the segment has proper time $\ell_t=\pi$.

Next, we extremize the spacelike segments that connect the bulk points $(\bm{u}_i, \bm{v}_i)$ with boundary points $(U_i,V_i)$. In the small $\delta$ limit, the length of the spacelike segment is
\begin{equation}
 \ell_s   =\log\!\left[
-\frac{16\,(V_{1}-\bm{v}_{1})(V_{1}-\bm{v}_{2})(V_{2}-\bm{v}_{1})(V_{2}-\bm{v}_{2})}
{V_{1}V_{2}\,(\bm{v}_{1}-\bm{v}_{2})^{2}\,\delta^{2}}
\right].
\end{equation}
Using \eqref{eqn:t_likecondBTZ} to eliminate $(\bm{u}_2,\bm{v}_2)$, for example, and then extremizing $\ell_s$ gives
\begin{equation}
\begin{split}
\frac{d \ell_s}{d\bm{u}_1} = \frac{ V_{1} + V_{2} + 2 V_{1} V_{2} \bm{u}_{1} - 2 \bm{v}_{1}
- (V_{1} + V_{2})\bm{u}_{1}\bm{v}_{1}
}{
(1 + V_{1}\bm{u}_{1})(1 + V_{2}\bm{u}_{1})(1 + \bm{u}_{1}\bm{v}_{1})
} = 0, \\
\frac{d \ell_s}{d\bm{v}_1} = \frac{ V_{1} + V_{2} + 2 V_{1} V_{2} \bm{u}_{1} - 2 \bm{v}_{1}
- (V_{1} + V_{2})\bm{u}_{1}\bm{v}_{1}
}{
(V_{1} - \bm{v}_{1})(-V_{2} + \bm{v}_{1})(1 + \bm{u}_{1}\bm{v}_{1})} = 0.
\end{split}
\end{equation}
As the numerators are identical, we have a one-parameter family of solutions, which we parametrize by $\bm{u}_1$,
\begin{equation}\label{eqn:s-likecondBTZ}
    \begin{split}
(\bm{u}_{1}, \bm{v}_{1}) &\;=\;
\left(
\bm{u}_{1}, \;
\frac{V_{1}+V_{2}+2 V_{1} V_{2}\,\bm{u}_{1}}
{2 + \bm{u}_{1}(V_{1}+V_{2})}
\right),
\\
(\bm{u}_{2}, \bm{v}_{2}) &\;=\;
\left(
-\frac{2+\bm{u}_{1}(V_{1}+V_{2})}{V_{1}+V_{2}+2 V_{1}V_{2}\,\bm{u}_{1}},\;
-\frac{1}{\bm{u}_{1}}
\right).
    \end{split}
\end{equation}
To determine where these points reside, we check the range of $s_1 \equiv \bm{u}_1 \bm{v}_1$ and $s_2 \equiv \bm{u}_2 \bm{v}_2$. From \eqref{eqn:s-likecondBTZ}, we see that $s_1 s_2 = 1$. However, only points satisfying $-1 \le s_1, s_2 \le 1$ lie within the BTZ geometry. The only allowed values are therefore $s_1 = s_2 = 1$, meaning the two endpoints are on the singularities. Imposing this condition fixes $\bm u_1^2= 1/(V_1 V_2)$. The time-ordered configuration is where $(\bm u_1, \bm v_1)$ lies on the future singularity, or $ \bm u_1 =1/\sqrt{V_1 V_2}$. This agrees with \cite{Doi:2022iyj, Doi:2023zaf}, which imposed that the endpoints were affixed to the singularities but without providing a justification. We therefore have
\begin{equation}
    \ell_s = \log{\frac{(V_1-V_2)^2}{V_1 V_2\, \delta^2}}.
\end{equation}
The total complex length $\ell_s + i\ell_t$ precisely reproduces the boundary result \eqref{eqn:Lbdy_timelike}. 

The Euclidean thermal two-point function on the cylinder parametrized by $(x,\tau)$ at inverse temperature $\beta$ is 
\begin{equation}
    \braket{0|\mathcal{O}(x,\tau) \mathcal{O}(0)|0} = \left( \frac{(\pi/\beta)^2}{\sinh\!{ \frac{\pi}{\beta} (x+i \tau)} \sinh\!{\frac{\pi}{\beta} (x-i\tau)}} \right)^\Delta.
\end{equation}
We analytically continue to the Lorentzian configuration $\tau = i(T_2-T_1),~x=0$, and the length extracted from the resulting correlator is equal to the quantity $\ell_s + i\ell_t$ we computed, upon substituting $V_i = e^{r_0 T_i}$ and $\beta = 2\pi/r_0$.
 
In Appendix \ref{sec:BTZB}, we show that in fact a continuous family of geodesics connects the bulk points we identified, all with the same total length $\ell$. One but not all of them lie entirely behind the horizon.

\section{Future directions}

Composite geodesics and their relationship with CFT correlators warrant a more systematic study. Computing CFT correlators in the WKB regime \cite{Balasubramanian:2012tu} or using worldline methods \cite{Maxfield:2017rkn} may lead to an unambiguous derivation of universal rules for computing composite geodesics. We expect this will clarify whether having a multiplicity of bulk geodesics is an intrinsic feature similar to the non-isometric encoding of infalling observers \cite{Akers:2022qdl}, or a sign the prescription needs further refinement. Studying these ideas in correlators of heavy operators \cite{Poland:2024rxv} may help identify technically natural CFT definitions of an observer and the bulk worldline on-shell action. A geodesic Witten diagram construction for timelike separations may be relevant here \cite{Hijano:2015zsa}. 

Our work and \cite{Doi:2022iyj, Doi:2023zaf} only studied examples in which the proper time $\ell_t$ was $\pi$. Setups with other values of $\ell_t$ would be useful for exploring $\ell_t$ as a manifestly diffeomorphism-invariant notion of bulk observer time. For related work on proper distance and time correlators in gravity, see \cite{Almheiri:2024xtw, Iliesiu:2024cnh, Carney:2024wnp, Sivaramakrishnan:2025srr, Giddings:2025xym}.

Our results suggest new ways to understand timelike entanglement entropy \cite{Doi:2022iyj, Doi:2023zaf, Milekhin:2025ycm}. In AdS$_3$/CFT$_2$ one may attempt to derive a bulk prescription by applying our methods to twist operators analytically continued to timelike separations. A possibly related connection between entropy and CFT correlators is the equivalence between the modular Hamiltonian and the stress-tensor OPE block \cite{Czech:2016xec, Czech:2016tqr, Haehl:2025ehf, Bub:2025lon}.

\begin{acknowledgments}
We thank Yanbei Chen, Sumit Das, Per Kraus, Jonathan Harper, David Poland, Mohamed Radwan, Gordon Rogelberg, and Alfred Shapere for discussions and comments on the draft. During this work, AS has been supported by the Heising-Simons Foundation ``Observational Signatures of Quantum Gravity'' collaboration grant 2021-2817; by the D.O.E., Office of High Energy Physics, under Award No. DE-SC0011632; and by the Walter Burke Institute for Theoretical Physics. HB is partially supported by the National Science Foundation grants NSF-PHY/211673 and NSF-PHY/2410647.
\end{acknowledgments}

\pagebreak
\bibliographystyle{utphys}
\bibliography{references}


\clearpage 
\pagebreak

\appendix 

\section{Geodesic method for global AdS$_3$} \label{sec:GlobalGeodesicMethod}

We derive \eqref{eqn:t-like_condition} using geodesic equations in AdS$_3$ for boundary points at the same angular location $(\Phi_1 = \Phi_2)$. We float the bulk endpoints and impose constraints to fix the timelike and spacelike segments. 

First we study the timelike segment. In global coordinates, the velocities obey
\begin{gather}
- (\cosh^2\!\rho)\, \dot{\tau}^2 + \dot{\rho}^2 + (\sinh^2\!\rho)\, \dot{\phi}^2 = -1,
\end{gather}
and the conserved quantities along a geodesic are
\begin{gather}\label{eqn:PtPphi}
(\cosh^2\!\rho)\,\dot{\tau}=P_\tau,\qquad (\sinh^2\!\rho)\,\dot{\phi}=P_\phi,
\end{gather}
with radial equation
\begin{gather}
\dot{\rho}^{\,2}=-1+P_\tau^2\,\text{sech}^2\!\rho - P_\phi^2\,\text{csch}^2\!\rho.
\end{gather}
Restricting to constant-$\phi$ solutions ($P_\phi=0$), the turning point $\rho_*$ obeys $P_\tau=\cosh\rho_*$. The proper time between endpoints $(\tau_1,\rho_1)$ and $(\tau_2,\rho_2)$ is
\begin{gather}
\ell_t
= \int d\lambda
= \Big\{\int_{\rho_1}^{\rho_*}+\int_{\rho_2}^{\rho_*}\Big\}
\frac{d\rho}{\sqrt{-1+P_\tau^2\,\text{sech}^2\!\rho}},
\end{gather}
which is maximized by $\rho_1=\rho_2=0$, yielding $\ell_t=\pi$. The entire timelike segment then sits at $\rho=0$ with $P_\tau=1$ and the endpoints satisfy $\tau_1-\tau_2=\pi$.

Next we study the spacelike segment, still imposing $P_\phi=0$. For the spacelike legs that connect $(\rho=0,\tau_{1,2})$ to boundary points $(\rho_\infty,T_{1,2})$,
\begin{gather}
- (\cosh^2\!\rho)\,\dot{\tau}^2 + \dot{\rho}^{\,2}=1,\qquad
c_\tau \equiv (\cosh^2\!\rho)\,\dot{\tau}
\end{gather}
Boundary conditions fix
\begin{gather}
\tau_1=\frac{\pi}{2}+\overline{T}_0,\quad
\tau_2=-\frac{\pi}{2}+\overline{T}_0,\\
c_\tau=\cot\!\left(\frac{T_1-T_2}{2}\right),\quad
\overline{T}_0=\frac{T_1+T_2}{2}.
\end{gather}
The regulated length of the two spacelike legs is
\begin{gather}
\ell_s
=\int_{0}^{\rho_\infty}\! \frac{2\ d\rho}{\sqrt{1+c_\tau^2\,\text{sech}^2\!\rho}}
=\;
\log\!\left[\frac{\sin^2\!\big(\tfrac{T_1-T_2}{2}\big)}{\delta^2}\right],
\end{gather}
with $\delta = e^{-\rho_\infty} \ll 1$.

Note that $\ell_t$ at $\rho =0$ is monotonically increasing in the separation between endpoints, as evident from \eqref{eqn:PtPphi} with $P_\tau = 1$ and $\rho =0.$

\section{Geodesic Calculation for BTZ}\label{sec:BTZB}
Here we rederive \eqref{eqn:t_likecondBTZ} using geodesic equations in the BTZ geometry. Let dots denote $d/d\lambda$ and primes $d/dv$. The nontrivial affine geodesic equations for \eqref{eqn:BTZmetric} with $\dot\phi=0$ are
\begin{equation}
\ddot u-\frac{2v}{1+uv}\,\dot u^{\,2}=0,
\qquad
\ddot v-\frac{2u}{1+uv}\,\dot v^{\,2}=0.    
\end{equation}
Parameterizing the curve as $u(v)$ gives
\begin{equation}
(1+uv)\,u''+2\big(u-vu'\big)u'=0,    
\end{equation}
whose general solution is
\begin{equation}\label{eqn:u(v)}
u(v)=\frac{a+bv}{1+av}\,,
\end{equation}
with real constants $a,b$. Along \eqref{eqn:u(v)}, the induced line element is
\begin{equation}
ds^2=-\frac{4}{(1+uv)^2}u'(v)\,dv^2
=\frac{4\,(a^{2}-b)\,dv^2}{\big(1+2av+bv^{2}\big)^{2}}.
\end{equation}
The curve is spacelike for $a^2-b>0$ and timelike for $a^2-b<0$. When $a^2>b$ there are two turning points,
\begin{equation}
    v_\pm=\frac{-a}{b}\pm\sqrt{\frac{a^2}{b}-1}\,,
\end{equation}
whereas for $a^2-b<0$, $v_\pm$ are complex and $u(v)$ is monotonic. For timelike separated bulk endpoints $(u_1,v_1)$ and $(u_2,v_2)$, define
\begin{equation}
y(v)=\frac{a+bv}{\sqrt{-a^2+b}}\ .    
\end{equation}
The proper time for this curve is then given as
\begin{align}
 \ell_t
&= \int_{v_1}^{v_2} dv\,\frac{2\sqrt{-a^2+b}}{1+2av+bv^2} = 2\tan^{-1}\!\left[\frac{y_2-y_1}{1+y_1y_2}\right], \label{eqn:t-lengthBTZ}
\end{align}
which is maximized when $1+y_1y_2=0$ with $\ell_t = \pi$. Using \eqref{eqn:u(v)}, this condition is equivalent to
\begin{equation}
v_2=-\frac{1+av_1}{a+bv_1}=-\frac{1}{u_1},
\quad
v_1=-\frac{1+av_2}{a+bv_2}=-\frac{1}{u_2},
\end{equation}
precisely matching \eqref{eqn:t_likecondBTZ}.

As a special case, we will consider the configuration with one endpoint on the future singularity. Choose $(u_1,v_1)=\big(1/v_+,\,v_+\big)$ with $v_+>0$ (so $u_1v_1=1$). From \eqref{eqn:u(v)},
\begin{eqnarray}
\frac{1}{v_+}=\frac{a+bv_+}{1+av_+}\quad\Rightarrow\quad b=\frac{1}{v_+^2}.    
\end{eqnarray}
The length is given by \eqref{eqn:t-lengthBTZ}, with endpoints fixed by $uv=1$. Let the corresponding $v$–values be $v_\pm$; then
\begin{eqnarray}
v_\pm=\pm\frac{1}{\sqrt{b}},\qquad y_+y_-=-1, \qquad
\ell_t =\pi.    
\end{eqnarray}
The timelike segment is therefore
\begin{equation}\label{eqn:t_geodesicBTZ}
u_t(v)=\frac{a+v/v_+^2}{1+av}, 
\end{equation}
with $v_+>0$ and $a \in (-1/v_+, 1/v_+)$. All such curves have the same length $i\pi$. Extremizing the spacelike legs fixes $v_+=\sqrt{V_1V_2}$ as also shown in Section \ref{BTZ}, while $a$ remains free within $|a|<1/v_+$. Only the curve with $a=0$ crosses the bifurcate horizon. For $a\neq 0$, the timelike curve $u_t(v)$ leaves the $u=0$ horizon at $(u,v)=(0,-a v_+^2)$, enters the right wedge, and re-enters the $v=0$ horizon at $(u,v)=(a,0)$. For future-directed curves, $a =0.$
\section{Proper time vs. modular time}
\label{modularobservers}

We have focused on reconstructing the proper time of bulk timelike geodesics. Other methods of reconstructing bulk time use modular flow \cite{Jafferis:2020ora, deBoer:2022zps, Leutheusser:2021frk} and are connected to geodesic time only in special cases. However, it is important to note that aside from the special case of Rindler, modular time is different from time experienced by an inertial observer. The former is a formal notion of time that exists for any state, while the latter is what appears more naturally in experiment and requires some well-defined clock. This distinction may be crucial for the holographic reconstruction of bulk time, and so we make this comparison explicit here.

In the right Rindler wedge of Minkowski space, the modular Hamiltonian is the boost generator. Under the conformal map that sends the wedge to the causal diamond $D(B_R)$ of a ball of radius $R$, the modular flow becomes the conformal Killing field, which in $(1+1)$ dimensions is
\begin{equation}
\label{eq:zetaB}
\zeta_B \;=\; \frac{\pi}{R}\Big[(R^2 - t^2 - x^2)\,\partial_t \;-\; 2 t x\,\partial_x\Big] .
\end{equation}
Integral curves $(t(\lambda),x(\lambda))$ of $\zeta_B$ obey
\begin{equation}
\label{eq:mod-ode}
\frac{dt}{dx} \;=\; \frac{-\,R^2 + t^2 + x^2}{2tx}\,,
\end{equation}
and can be written equivalently as the one–parameter family of hyperbolae labeled by a constant $\xi_0$,
\begin{equation}
\label{eq:mod-traj}
\frac{t^2 - (R+x)^2}{t^2 - (R-x)^2} \;=\; \frac{\xi_0^{\,2}}{4R^{2}}\,,
\end{equation}
which furnishes a convenient check of \eqref{eq:mod-ode}. Note that $dx/dt$ is not constant, indicating the trajectory is non-inertial. The relation between modular time $\eta$ and $t$ is
\begin{equation}
\label{eq:mod-time}
\eta(t,x) \;=\; \tanh^{-1}\!\left(\frac{2Rt}{R^2 + t^2 - x^2}\right).
\end{equation}
It is easy to check that for $x=0$, the observer is inertial, and furthermore that modular time is non-linearly related to proper time. Modular time is therefore not directly measurable by the clock of some observer in this setup.

\end{document}